\documentclass[twocolumn,showpacs]{revtex4}

\usepackage{graphicx}
\usepackage{dcolumn}
\usepackage{bm}
\newcommand{\beq}{\begin{equation}}
\newcommand{\eeq}{\end{equation}}
\newcommand{\bea}{\begin{eqnarray}}
\newcommand{\eea}{\end{eqnarray}}
\newcommand{\bec}{\begin{center}}
\newcommand{\enc}{\end{center}}
\newcommand{\alp}{\alpha}
\newcommand{\om}{\omega}
\newcommand{\Om}{\Omega}
\newcommand{\eps}{\epsilon}
\newcommand{\rmd}{{\rm d}}
\newcommand{\rmg}{{\rm g}}
\newcommand{\rmi}{{\rm i}}
\newcommand{\rmx}{{\rm x}}

\newcommand{\tiD}{\tilde{\Delta}}

\newcommand{\veck}{\bm{k}}

\begin{document}
\title{Quantum Zeno and anti-Zeno effects 
by indirect measurement with finite errors}
\author{Kazuki Koshino}
\email{ikuzak@postman.riken.go.jp}
\affiliation{
Frontier Research System, 
The institute of Physical and Chemical Research (RIKEN),
Hirosawa 2-1, Wako, Saitama 351-0198, Japan}
\author{Akira Shimizu}
\affiliation{
Department of Basic Science, University of Tokyo, 
3-8-1 Komaba, Tokyo 153-8902, Japan}
\date{\today}
\begin{abstract}
We study the quantum Zeno effect and the anti-Zeno effect in the case 
of `indirect' measurements, where a measuring apparatus does not act 
directly on an unstable system, for a realistic model with finite 
errors in the measurement.  A general and simple formula for the 
decay rate of the unstable system under measurement is derived.
In the case of 
a Lorentzian form factor, we calculate the full time evolutions 
of the decay rate, the response of the measuring apparatus, and the 
probability of errors in the measurement.  It is shown that 
not only the response time but also the detection efficiency plays a 
crucial role.  We present the prescription for observing the quantum 
Zeno and anti-Zeno effects, as well as the prescriptions for avoiding 
or calibrating these effects in general experiments.
\end{abstract}
\maketitle


It was predicted in a classic paper \cite{classic1}
that repeated measurements on 
a quantum unstable system, at time intervals $\tau_{\rm m}$, 
suppress the decay of the system for small $\tau_{\rm m}$ ---
the so-called quantum Zeno effect (QZE) .
It was assumed there that 
each measurement is completely ideal, i.e., 
it takes only an infinitesimal time, 
there is no error in the measurement, 
and the post-measurement state is exactly given by the 
projection 
postulate. 
However, any physical experiments do not satisfy all of these assumptions,
hence more careful studies have been desired.

According to the general measurement theory, 
not only the unstable system in question but also 
a part of the measuring apparatus should be treated as 
a quantum system subject to the Schr\"odinger equation
\cite{laser_th0,classic2,classic3,laser_th1}.
It was clarified by such theories that the response time 
$\tau_{\rm r}$ of the apparatus 
corresponds 
to $\tau_{\rm m}$ of Ref.\ \cite{classic1}.
However, effects of the errors in the measurement are yet to be 
explored, because
the probability $\varepsilon$ of getting an erroneous 
result is determined not only by a finite response time $\tau_{\rm r}$,
but also by the detection efficiency $1 - \varepsilon_\infty$
(i.e., the apparatus occasionally fails to detect the decay 
even after an infinitely long waiting time). 
%
Moreover, these pioneering theories,
as well as pioneering experiments \cite{laser_exp1,laser_exp2,AZE_exp1}, 
studied the case of
`direct' measurements,
where the apparatus acts directly on the unstable system
(e.g., shines laser light to excited atoms). 
In such a case, however, 
the dynamics of the unstable system would be affected 
by the apparatus even if the unstable system were 
a classical system, and thus
the Zeno effect in direct measurements might not be 
peculiar to quantum systems.
Hence, the most interesting case of `indirect' measurements
is yet to be explored, 
where the apparatus 
does not act directly on the unstable system, 
but detects a signal mediated by some field. 
A promising theory of an indirect measurement
was developed by Schulman \cite{nonlaser_th1}.
However, since his model was an abstract one, 
application to real physical systems is not straightforward. 
For example, 
it did not clarify the conditions to observe 
{\em acceleration} of the decay by measurements, 
the anti-Zeno effect (AZE), 
which has also been attracting much attention 
\cite{AZE_exp1,AZE_th1,AZE_th3}.
To apply real experiments on QZE and AZE, 
more realistic models should be analyzed.
Such analysis should also 
be important to {\em general} experiments, 
because the QZE or AZE might slip 
in advanced experiments in the near future.
The purpose of this paper is to 
present a theory that satisfies all these requirements, 
using a realistic model of an indirect measurement 
with finite errors.


The total system in our model is composed of three parts,
(i) an unstable two-level system, 
which is initially in the excited state 
$|\rmx\rangle$ with the transition energy $\Om_0$ to 
the ground state $|\rmg\rangle$,
(ii) a field whose eigenmodes are labeled by a wavevector 
$\veck$ with any dimension, 
a quantum of which is emitted by the unstable system when it decays to 
$|\rmg\rangle$,
and (iii) a measuring apparatus that detects the emitted quantum
by absorbing it, from which an 
observer gets to know the decay of the unstable system.
A typical example for (i) 
is an excited atom, 
which emits a photon 
upon decay, and the photon is 
detected by a photodetector 
such as a photomultiplier.
Therefore, we hereafter call (i), (ii) and (iii) 
as an `atom', `photon' and `detector,' respectively, 
although the theory is applicable to other systems as well.
In this model, 
neither a projection operator 
nor the interaction Hamiltonian of the detector
acts on the atom.
Moreover, the measurement is of negative-result type, 
where no signal is detected until the atom decays: 
The QZE or AZE occurs just by waiting for the decay.

The role of a detector is to convert a photon 
into other kinds of elementary excitations,
which finally yield macroscopic signals after magnification processes,
usually obeying classical mechanics.
As a model of (the relevant part of) the detector, 
we assume elementary excitations (e.g, electron-hole pairs)
with a continuous spectrum, 
into which photons are converted. 
By taking the energy of $|\rmg\rangle$ zero, 
the Hamiltonian of the system is taken as follows
(with $\hbar = 1$);
\bea
{\cal H}&=&\Om_0 |\rmx\rangle\langle\rmx|+{\cal H}_1+{\cal H}_2,
\label{eq:H}
\\
{\cal H}_1 \! &=& \! 
\int\rmd\veck
\left[
\left(g_{\veck}|\rmx\rangle\langle\rmg|b_{\veck}+{\rm H.c.}\right)
+\eps_{\veck}\ b_{\veck}^{\dagger}b_{\veck}
\right],
\label{eq:H1}
\\
{\cal H}_2 \! &=& \! 
\int \! \! \int \! \rmd\veck \, \rmd\om 
\left[
\left(\zeta_{\veck\om}b_{\veck}^{\dagger}c_{\veck\om}+{\rm H.c.}\right)
+\om c_{\veck\om}^{\dagger}c_{\veck\om}
\right].
\label{eq:H2}
\eea
Here, 
$b_{\veck}$ is the annihilation operator 
for a photon with energy $\eps_{\veck}$,
$g_{\veck}$ represents the atom-photon coupling.
The photonic dispersion relation can be nonlinear 
($\eps_{\veck} \not \propto |\veck|$) as in, 
say, photonic crystals.
Every photon mode is linearly coupled to a continuum of 
bosonic \cite{bosonic} elementary excitations, denoted by $c_{\veck\om}$, 
in the detector, with the coupling constant $\zeta_{\veck\om}$.
The commutation relations are normalized as 
$[b_{\veck},b^{\dagger}_{\veck'}]=\delta(\veck-\veck')$ and
$[c_{\veck\om},c^{\dagger}_{\veck'\om'}]=\delta(\veck-\veck')\delta(\om-\om')$.
The lifetime $\tau_{\veck}$ of a photon
is determined by $\zeta_{\veck\om}$.
We will show later that $\tau_{\veck} \simeq \tau_{\rm r}$,  
the response time of the detector.
In most experiments, 
the detector does not cover the whole solid angle around the atom.
When a photon is emitted in the uncovered direction, it cannot be detected
and has a long lifetime.
As we will demonstrate later, we can encompass such realistic situations
by allowing $\veck$-dependence of $\tau_{\veck}$.
We here put $\zeta_{\veck\om}=\sqrt{\eta_{\veck}}$ \cite{omegadep},
which results in $\tau_{\veck}=(2\pi\eta_{\veck})^{-1}$.


The photon-detector 
part, ${\cal H}_1+{\cal H}_2$, can be diagonalized
in terms of the coupled-mode operator \cite{Fano}, which 
in this case is given by
$ 
B_{\veck\mu}=\alp_{\veck}(\mu)b_{\veck}+\int\rmd\om\ \beta_{\veck}(\mu,\om)c_{\veck\om},
$ 
where
$
\alp_{\veck}(\mu)=\sqrt{\eta_{\veck}}/(\mu-\eps_{\veck}+\rmi\pi\eta_{\veck})
$
and
$
\beta_{\veck}(\mu,\om)=\eta_{\veck}/(\mu-\eps_{\veck}+\rmi\pi\eta_{\veck})(\mu-\om+\rmi\delta)
+\delta(\mu-\om)
$.
The commutation relations for $B_{\veck\mu}$ is given 
by $[B_{\veck\mu},B^{\dagger}_{\veck'\mu'}]=\delta(\veck-\veck')\delta(\om-\om')$.
Inversely, $b_{\veck}$ is expressed in terms of $B_{\veck\mu}$ as
$b_{\veck}=\int\rmd\mu\ \alp^{\ast}_{\veck}(\mu)B_{\veck\mu}$.
The Hamiltonian ${\cal H}$ can then be rewritten as follows,
\bea
{\cal H}&=&\Om_0 |\rmx\rangle\langle\rmx|
+\int\! \! \int\rmd\veck\rmd\mu \ 
\mu B_{\veck\mu}^{\dagger}B_{\veck\mu}\nonumber\\
&+& \! \int \! \! \int\rmd\veck \, \rmd\mu
\left[\frac{\sqrt{\eta_{\veck}}g_{\veck}}
{\mu-\eps_{\veck}-\rmi\pi\eta_{\veck}}|\rmx\rangle\langle\rmg|B_{\veck\mu}+{\rm H.c.}
\right].
\label{eq:re1}
\eea
We further rewrite ${\cal H}$ using $\bar{B}_{\mu}$ that is defined by
\beq
\bar{B}_{\mu}=\frac{1}{\bar{g}_{\mu}}\int\rmd\veck
\frac{\sqrt{\eta_{\veck}}g_{\veck}}{\mu-\eps_{\veck}-\rmi\pi\eta_{\veck}}B_{\veck\mu},
\eeq
where $\bar{g}_{\mu}$ satisfies
\beq
|\bar{g}_{\mu}|^2=\int\rmd\veck\left|\frac{\sqrt{\eta_{\veck}}g_{\veck}}
{\mu-\eps_{\veck}-\rmi\pi\eta_{\veck}}\right|^2,
\label{eq:g2}
\eeq
so that $\bar{B}_{\mu}$ is normalized as 
$[\bar{B}_{\mu},\bar{B}^{\dagger}_{\mu'}]=\delta(\mu-\mu')$.
Then, the Hamiltonian finally reduces to
\bea
{\cal H}&=&\Om_0 |\rmx\rangle\langle\rmx|+\bar{\cal H}_1+\bar{\cal H}_2,
\label{eq:renH}
\\
\bar{\cal H}_1&=&
\int\rmd\mu \left[
\left(\bar{g}_{\mu}|\rmx\rangle\langle\rmg|\bar{B}_{\mu}+
{\rm H.c.}\right)
+ \mu \bar{B}_{\mu}^{\dagger}\bar{B}_{\mu}
\right],
\label{eq:renH1}
\eea
where $\bar{\cal H}_2$ 
consists of coupled-modes which do not interact with the atom.
Now the physical meaning of $\bar{B}_{\mu}$ becomes apparent; it
is part of the coupled-modes with energy $\mu$ that interact with
the atom, whereas the other part is isolated.
In terms of such operators, 
the Hamiltonian is expressed in the renormalized form, 
Eqs.~(\ref{eq:renH}) and (\ref{eq:renH1}),
where the atom is coupled to a single continuum of $\bar{B}_{\mu}$
with a coupling constant $\bar{g}_{\mu}$,
which is called the form factor of interaction.
The form factor under measurement
is determined by Eq.~(\ref{eq:g2}),
from $\eps_{\veck}$, $g_{\veck}$, and $\eta_{\veck}$.
It should be noted that there exists a sum rule 
$\int\rmd\mu|\bar{g}_{\mu}|^2=\int\rmd\veck|g_{\veck}|^2$,
which holds for any functional form of $\eta_{\veck}$.

We first estimate the decay rate $\Gamma$ under the measurement
by a lowest-order perturbation in $g_{\veck}$.
We note that the straightforward application of Fermi's golden rule 
using ${\cal H}_1$ as
the interaction term gives a wrong result, 
because strong effects of the detector are not involved.
It is essential to use
the renormalized form $\bar{\cal H}_1$ 
as the interaction term.
We then obtain a 
simple formula for 
the decay rate under the measurement;
\beq
\Gamma= 
2\pi\int\rmd\veck\left|\frac{\sqrt{\eta_{\veck}}g_{\veck}}
{\Om_0-\eps_{\veck}-\rmi\pi\eta_{\veck}}\right|^2,
\label{eq:RDR}
\eeq
which should be compared with the free decay rate,
$
\Gamma_0=
2\pi\int\rmd\veck |g_{\veck}|^2 \delta(\Om_0-\eps_{\veck})
$.
Note that Eq.\ (\ref{eq:RDR}) includes non-perturbative effects of
$\eta_{\veck}$.
The formula clearly shows that 
the most important effect of the measurement (i.e., of finite $\eta_{\veck}$) 
is to renormalize the form factor $\bar{g}_{\mu}$,
and that the QZE (or AZE) occurs through the renormalization.
Note that the formula is general,
which holds for any forms of $\eps_{\veck}$, $g_{\veck}$ and $\eta_{\veck}$,
and for any dimension of $\veck$.
It is applicable, not only to 
spontaneous decay of an atom, but also to many other unstable systems
if their Hamiltonian can be approximated by 
Eqs.~(\ref{eq:H})-(\ref{eq:H2}) \cite{bosonic}.
Moreover, the formula is also applicable to 
the case where the detector does not cover the full solid angle, 
yielding the detection efficiency $< 1$.
In fact, suppose that 
only photons which are 
emitted in some solid angle ${\rm S_d}$ in the three-dimensional 
space are coupled to 
the detector, i.e., 
$\eta_{\veck}=0$ for $(\theta,\phi)\notin {\rm S_d}$, where
$\veck=(k\sin\theta\cos\phi,k\sin\theta\sin\phi,k\cos\theta)$.
Then, Eq.\ (\ref{eq:RDR}) yields the simple formula;
\begin{eqnarray}
\Gamma &=&
2\pi \int k^2{\rm d}k \int_{{\rm S}\in{\rm S_d}} {\rm dS}
\left|\frac{\sqrt{\eta_{\veck}}g_{\veck}}
{\Om_0-\eps_{\veck}-\rmi\pi\eta_{\veck}}\right|^2
\nonumber\\
&+&
2\pi \int k^2{\rm d}k \int_{{\rm S} \notin{\rm S_d}} {\rm dS}
\left|g_{\veck}\right|^2 \delta(\Om_0-\eps_{\veck}),
\label{eq:RDRpart}
\end{eqnarray}
where ${\rm d S} = \rmd\cos\theta\rmd\phi$.


Now we embody the above general results in an 
example in three dimension, 
in which 
(i) $\eta_{\veck}=\eta \equiv (2 \pi \tau)^{-1}$ 
for $(\theta,\phi)\in {\rm S_d}$,
whereas $\eta_{\veck}=0$ for $(\theta,\phi)\notin {\rm S_d}$,
(ii) $\eps_{\veck}=k$ (we take $c=1$), 
and (iii) 
$g_{\veck}$ takes the Lorentzian form after the angular integration;
$
k^2 \int_{{\rm S} \in {\rm S_d}}{\rm dS}\ |g_{\veck}|^2
= (1-\varepsilon_{\infty})\gamma\Delta^2/[(k-k_0)^2+\Delta^2]
$, and
$
k^2 \int_{{\rm S} \notin {\rm S_d}}{\rm dS}\ |g_{\veck}|^2
=\varepsilon_{\infty}\gamma\Delta^2/[(k-k_0)^2+\Delta^2]
$, where $k_0\gg\Delta$.
In a case where $g_{\veck}$ depends only on $k$,
$\varepsilon_{\infty}$ is simply given by $\varepsilon_{\infty}=1-|{\rm S_d}|/4 \pi$.
In general cases, however, 
$\varepsilon_{\infty}$ depends on both ${\rm S_d}$ and $g_{\veck}$.
It will turn out that $1-\varepsilon_{\infty}$ corresponds to the
detection efficiency, i.e., 
the probability of errors approaches $\varepsilon_{\infty}$
as $t \to \infty$.
The meanings of $\gamma$ and $\Delta$ become transparent for $\Om_0=k_0$,
i.e., when the atomic transition energy coincides with the center of the Lorentzian.
For $\Delta\gg\gamma$, the decay rate of the atom is given by $2\pi\gamma$,
while the transition from the initial quadratic decrease 
to the exponential decrease in the survival probability
occurs at $t \simeq \tau_{\rm j} \equiv 2/\Delta$, 
which is called the `jump time' 
\cite{nonlaser_th1,AZE_th3}.
Using Eq.~(\ref{eq:RDRpart}), 
the renormalized lowest-order decay rate is evaluated as
\beq
\frac{2\pi\gamma (1-\varepsilon_{\infty}) \Delta\tiD}{(\Om_0-k_0)^2+\tiD^2}+
\frac{2\pi\gamma \varepsilon_{\infty} \Delta^2}{(\Om_0-k_0)^2+\Delta^2}
\equiv
\Gamma(\eta,\varepsilon_{\infty}),
\label{eq:RDR_L}\eeq
where $\tiD=\Delta+\pi\eta$. 
This indicates that
the effect of measurement on the decay dynamics become significant 
only for large $\eta$ satisfying 
$\eta\gtrsim\Delta$,
i.e., $\tau_{\rm j} \gtrsim (2\pi\eta)^{-1} = \tau \simeq \tau_{\rm r}$,
in accordance with the pervious studies \cite{laser_th1,nonlaser_th1}.

\begin{figure}
\includegraphics{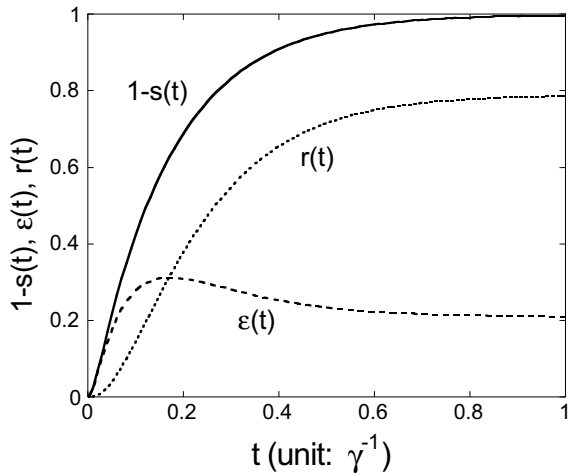}
\caption{\label{fig:ABC}
The decay probability $1-s(t)$ (solid curve), 
the probability of getting an erroneous result $\varepsilon(t)$ (broken curve), 
and the probability of getting a detector response $r(t)$ (dotted curve),
when $\Om_0-k_0=0$, $\Delta=100\gamma$, $\eta=2\gamma$, and 
$\varepsilon_{\infty}=0.2$.
}\end{figure}

To see what is going on, 
we now calculate 
the temporal evolution of the wavefunction 
from the initial state $|\rmx,0,0\rangle$.
This can be pursued analytically for the Lorentzian form factor.
By putting
$|\psi(t)\rangle=e^{-\rmi{\cal H}t}|\rmx,0,0\rangle
=f(t)|\rmx,0,0\rangle+\int\rmd\veck\ f_{\veck}(t)|\rmg,\veck,0\rangle
+\int\int\rmd\veck\rmd\om\ f_{\veck\om}(t)|\rmg,0,\veck\om\rangle$,
we calculate three probabilities;
$s(t)=|f(t)|^2$ (survival probability of the atom),
$\varepsilon(t)=\int\rmd \veck|f_{\veck}(t)|^2$ 
(probability that the atom has decayed but the emitted photon 
is not absorbed by the detector), and
$r(t)=\int\int\rmd \veck\rmd\om|f_{\veck\om}(t)|^2$ 
(probability that the emitted photon is absorbed).
Assuming fast classical magnification processes, we 
can interpret $r(t)$ as the probability 
of getting a detector response,
whereas $\varepsilon(t)$ is the probability 
that the detector reports an erroneous result.
One of the advantages of the present theory is that all of 
these interesting quantities can be calculated.
For example, $s(t)$ is given by
$
s(t)=4\pi^2|c_{123}e^{-\rmi\om_1t}+c_{231}e^{-\rmi\om_2t}+c_{312}e^{-\rmi\om_3t}|^2,
$
where $c_{ijk}$ is given by
$
c_{ijk}=\gamma\Delta
[(\Delta+p\pi\eta)(\om_i-k_0)^2+(\tiD-p\pi\eta)\Delta\tiD]
[(\om_i-\om_i^{\ast})(\om_i-\om_j)(\om_i-\om_j^{\ast})(\om_i-\om_k)(\om_i-\om_k^{\ast})]^{-1},
$
and $\om_j$ ($j=1,2,3$) are the solutions of the cubic equation,
$
(\om_j-\Om_0)(\om_j-k_0+\rmi\Delta)(\om_j-k_0+\rmi\tiD)=
\pi\gamma\Delta[\om_j-k_0+\rmi(p\Delta+(1-p)\tiD)].
$
Figure \ref{fig:ABC} plots $1-s(t), \varepsilon(t), r(t)$
for $\eta=2\gamma$.
At the initial time stage ($t\ll\tau=(2\pi\eta)^{-1}$),
$\varepsilon(t)$ increases almost in parallel with $1-s(t)$,
and the detection probability $r(t)[=1-s(t)-\varepsilon(t)]$ 
remains almost zero.
Around $t\sim\tau$, 
the emitted photon is gradually absorbed and $r(t)$ starts to rise. 
Hence, the photon lifetime
$\tau$ can be regarded as the response time $\tau_{\rm r}$ of the detector.
The probabilities finally approaches the asymptotic values,
$1-s(t)\rightarrow 1$, $\varepsilon(t)\rightarrow\varepsilon_{\infty}$, 
and $r(t)\rightarrow 1-\varepsilon_{\infty}$.

\begin{figure}
\includegraphics{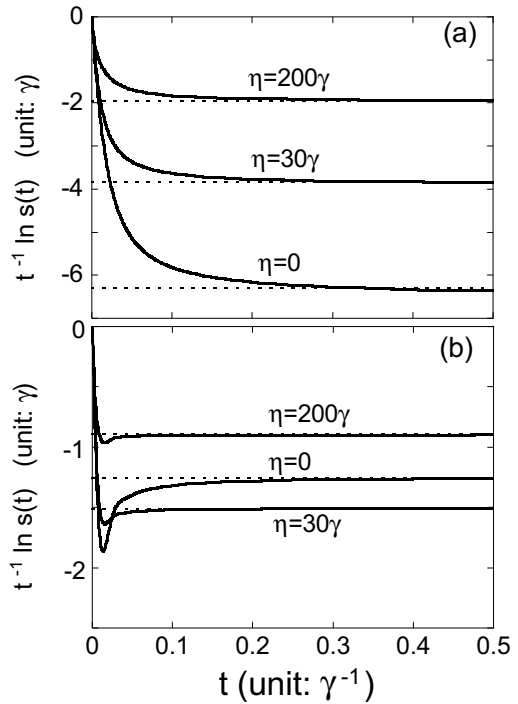}
\caption{\label{fig:logA}
Plots of $t^{-1}\ln s(t)$. $\Delta=100\gamma$ and 
$\varepsilon_{\infty}=0.2$ in both figures.
$\Om_0-k_0$ is taken $0$ ($<\Delta$) in (a), and 200$\gamma$ ($>\Delta$) in (b).
The dotted lines show the renormalized lowest-order decay rate, 
Eq.~(\ref{eq:RDR_L}). 
}\end{figure}

To see the temporal behavior of  $s(t)$ more clearly,
we have plotted $t^{-1}\ln s(t)$ as a function of time in Fig.~\ref{fig:logA},
for several different values of the parameters.
At the beginning of the decay ($t\lesssim\tau_{\rm j}$), 
$t^{-1}\ln s(t)$ decreases linearly as 
$t^{-1}\ln s(t)=-(\int\rmd\veck|g_{\veck}|^2)t=-\pi\gamma\Delta t$,
for any value of $\Om_0-k_0$ and for any values of 
the detector parameters $\eta$ and $\varepsilon_{\infty}$.
%
Then, for $t \gtrsim \tau_{\rm j}$,
$t^{-1}\ln s(t)$ approaches a constant value,
which is well approximated by 
$\Gamma(\eta,\varepsilon_{\infty})$
(dotted lines).
These plots demonstrate that
the decay dynamics is well described, except for the initial deviation, 
by the exponential decay with the renormalized lowest-order decay rate.
%
In fact, 
we can show analytically that formula (\ref{eq:RDR_L}) is 
a good approximation to the asymptotic decay rate if 
$\gamma \ll |\Om_0-k_0+\rmi\Delta|^2/\Delta$.

\begin{figure}
\includegraphics{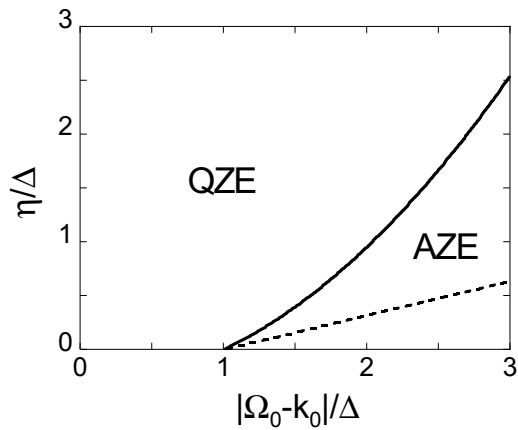}
\caption{\label{fig:PD}
The phase diagram for the QZE and the AZE for a case of Lorentzian form factor
(valid for any $\varepsilon_{\infty}$).
The solid curve divides the QZE region and the AZE region.
The dotted line shows the value of $\eta$ at which 
the decay rate is maximized for each value of $|\Om_0-k_0|$.
}\end{figure}

There is a remarkable difference between Figs.~\ref{fig:logA}(a) and (b),
where different values of $|\Om_0-k_0|/\Delta$ are employed.
In case of $|\Om_0-k_0|/\Delta=0$ [see Fig.~\ref{fig:logA}(a)],
the decay rate decreases monotonously by increasing $\eta$, 
i.e., the QZE occurs at any value of $\eta$.
Contrarily, in case of  $|\Om_0-k_0|/\Delta=2$ [Fig.~\ref{fig:logA}(b)],
the decay is enhanced for small $\eta$ ($=30\gamma$), 
while it is suppressed for large $\eta$ ($=200\gamma$).
Thus, the AZE takes place for small $\eta$. 
By analyzing Eq.~(\ref{eq:RDR_L})
as a function of $|\Om_0-k_0|$ and $\eta$,
the `phase diagram' discriminating the QZE and AZE is generated,
which is shown in Fig.~\ref{fig:PD}.
The `phase boundary' (solid curve) is given by
\beq
\eta^{\rm (b)}=
[(\Om_0-k_0)^2-\Delta^2]/\pi\Delta,
\label{eq:pb}\eeq 
on which the decay rate is not altered from the free 
rate $\Gamma_0$,
while the decay rate takes the maximum value,
\beq
\frac{\Gamma(\eta^{\rm (m)},\varepsilon_{\infty})}{\Gamma_0}
=\varepsilon_{\infty}+(1-\varepsilon_{\infty})
\left(\frac{|\Om_0-k_0|^2+\Delta^2}{2\Delta|\Om_0-k_0|}-1\right),
\eeq
on the dotted line, which is given by
\beq
\eta^{\rm (m)}=
[|\Om_0-k_0|-\Delta]/\pi.
\eeq
We find that $\eta^{\rm (b)}$ and $\eta^{\rm (m)}$ 
do not depend on
$\varepsilon_{\infty}$.


We finally discuss the significance of our results, 
for experiments on the QZE or AZE, and for {\em general} experiments.
As mentioned earlier,
indirect measurements are necessary, which have not been performed yet,
for the complete experimental verification of the QZE and AZE.
For such experiments, 
formula (\ref{eq:RDR}) gives the necessary condition: 
$\Gamma$ of Eq.\ (\ref{eq:RDR})
should significantly differ from the free decay rate $\Gamma_0$.
In the Lorentzian case, this can be decomposed into the 
following conditions:
(i) $\tau_{\rm r}$ should be short enough; 
$\tau_{\rm r}\lesssim\tau_{\rm j}$
(which is a well-known condition),
(ii) $\eta$ should not be close to the phase boundary 
(\ref{eq:pb}),
and (iii) $\varepsilon_{\infty}$ should be so small that
the first term of Eq.\ (\ref{eq:RDR_L}) becomes dominant.
Moreover, the QZE or AZE should be chosen according to 
the phase diagram, Fig.~\ref{fig:PD}:
e.g., the AZE is most detectable on the dotted line.
%
%
On the other hand, in general experiments,
one usually wants to {\em avoid} the QZE and AZE
in order to get correct results.
Considering recent rapid progress of experimental techniques
and diversification of experimental objects, 
we expect that the QZE or AZE would slip
in advanced experiments in the near future.
To avoid the QZE and AZE, 
one must design the experimental setup to break at least 
one of the above conditions.
For example, 
when performing an experiment with a high 
time resolution such that $\tau_{\rm r} \lesssim \tau_{\rm j}$, 
then Eqs.\ (\ref{eq:RDRpart}) and (\ref{eq:RDR_L}) suggest that
$\varepsilon_{\infty}$ should be {\em increased}.
If $\varepsilon_{\infty}$ cannot be increased to keep the sensitivity of such high-speed
measurement,  
then one should adjust parameters in such a way that
Eq.\ (\ref{eq:pb}) is satisfied, or, 
one should calibrate the observed value using
our results, such as Eqs.\ (\ref{eq:RDRpart}) and (\ref{eq:RDR_L}),
to obtain the free decay rate.



\begin{thebibliography}{99}
\bibitem{bosonic}
In general, excitation of fermion pairs at low density are well described by
a collection of harmonic oscillators.
\bibitem{omegadep}
It is straightforward to include $\omega$ dependence 
of $\zeta_{\veck\om}$. 
We find that it just introduces uninteresting complexity.
\end{thebibliography}
\end{document}